# Stochastic resolution of identity to CC2 for large systems: ground-state properties

Chongxiao Zhao[1,2,*] and Wenjie Dou,[1,2,*]

[1]*Department of Chemistry, School of Science, Westlake University, Hangzhou, Zhejiang 310024, China*

[2]*Institute of Natural Sciences, Westlake Institute for Advanced Study, Hangzhou, Zhejiang 310024, China*

*Email: zhaochongxiao@westlake.edu.cn; douwenjie@westlake.edu.cn*

A stochastic resolution of identity approach (sRI) is applied to the second-order coupled cluster singles and doubles (CC2) model to calculate the ground-state energy. Utilizing a set of stochastic orbitals to optimize the expensive tensor contraction steps in CC2, we greatly reduce the overall computational cost. Compared with the RI-CC2 model, the sRI-CC2 achieves scaling reduction from $O(N^5)$ to $O(N^3)$, where $N$ is a measure for the system size. When applying the sRI-CC2 to a series of hydrogen dimer chains, we demonstrate that the sRI-CC2 accurately reproduces RI-CC2 results for the correlation energies and exhibits a scaling of $O(N_H^{2.71})$, with $N_H$ being the number of hydrogen atoms. Our calculations with different systems and basis sets show small changes in standard deviations, which indicates a broad applicability of our approach to various systems.

## I. INTRODUCTION

Developing accurate and affordable electronic structure theory for complex systems is still one of the most challenging problems in theoretical chemistry. Among the available methodologies, the coupled cluster singles and doubles model (CCSD) has been proved to be a valuable one, which scales as $O(N^6)$ (with N being a measure of the system size). The CCSD model was first implemented by Purvis and Bartlett[1] in 1982 and became popular in realistic electronic structure calculations due to further development by Koch *et al*.[2,3] Later in 1995, Christiansen and Koch *et al*.[4] reported the formulation and implementation of the CC2 model, which scales as $O(N^5)$, as an approximation to CCSD. Significantly, the CC2 model provides the ground state energy as well as the excitation energy from the single excitation dominant transition, which is correct to the second order in the fluctuation operator. Between 2000 and 2001, Hald *et al*.[5-7] took advantage of integral-direct approach to handle the 4-index electron repulsion integrals (ERIs), which reduced the computational cost of the CC2 model and extended its applicability to larger molecular systems. In addition to the integral-direct approach, the introduction of the resolution-of-the-identity (RI) approximation[8,9] by Hättig and Weigend[10] in 2000 greatly ameliorated the bottlenecks of CPU time and storage and has enabled a widespread use of the CC2 model.[11]

There have been many successful cases[12-17] of the ground state energy calculations by the RI approach. However, the RI approximation still scales as $O(N^5)$, consuming huge memory and high disk space. In addition, the CC2 calculation of the ground state properties needs to be achieved through an iterative process, which makes it obviously inferior to other methods such as MP2 in time consumption. These factors weaken the advantages of CC2 and make people pay more attention to its performance in the calculation of the electronic excited state energy.[18]

For these reasons, a stochastic orbital approach is introduced to the RI approximation, abbreviated as sRI approach, to further reduce the scaling of the CC2 ground state energy calculations. The sRI approximation has been formulated and implemented for a variety of electronic structure theory, including MP2[19-21], DFT[22,23], GW[24] *et al*. In the sRI approach, a set of random orbitals are introduced to simplify the achievement of 4-index ERIs and lower the rank of computational costs from $O(N^5)$ to $O(N^3)$, when basically maintains the original accuracy. In the subsequent research[25-27], this sRI approach has been further applied to the second-order Matsubara Green's function (sRI-GF2) theory and turned out to be a practical approach for large weakly-correlated systems. Inspired by its high performance, we introduce the sRI method to the calculation of the CC2 ground state energy.

In the paper, we develop the sRI-CC2 model to further reduce the scaling. We apply the sRI approximation to the CC2 theory for the ground state properties and show that the sRI approximation reduces the scaling of CC2 from $O(N^5)$ to $O(N^3)$. We test the performance of sRI-CC2 for hydrogen dimer chains as well as a list of molecular systems. We show that sRI-CC2 reproduce the results from RI-CC2 in Q-Chem package[28] for intensive properties, with a stochastic error that does not depend on system size. We further analyze the error to demonstrate the applicability of the sRI-CC2 to a variety of molecular systems.

This paper is organized as follows: In the section II, we briefly review RI and sRI methods. We also demonstrate the detailed implementation of sRI-CC2 in this section. In the section III, a comparison of RI-CC2 and sRI-CC2 approaches for a series of molecules and basis sets is presented, with emphasis on the scaling of CPU time and the assessment of the correlation energies and standard deviations. Finally, we conclude in the section IV.

## II. THEORY

### A. Notation

We use the notations in Table I to represent the items used in the main text. In particular, the total number of AO basis functions, auxiliary basis functions, occupied, and unoccupied sets of MOs are denoted as $N_{AO}$, $N_{aux}$, $N_{occ}$, and $N_{virt}$ respectively.

TABLE I. Summary of notations in the following equations.

| item | function or indices |
|---|---|
| AO Gaussian basis functions | $\chi_\alpha(r_1), \chi_\beta(r_1), \chi_\gamma(r_1), \chi_\delta(r_1), \ldots$ |
| auxiliary basis functions | $P, Q, R, S, \ldots$ |
| general sets of AOs | $\alpha, \beta, \gamma, \delta, \ldots$ |
| general sets of MOs | $p, q, r, s, \ldots$ |
| occupied (active) MOs | $i, j, k, l, \ldots$ |
| unoccupied (virtual) MOs | $a, b, c, d, \ldots$ |

### B. Resolution of identity (RI) and stochastic resolution of identity (sRI)

Before introducing the stochastic resolution of identity, we briefly review the resolution of identity approach. In the RI approximation, the 4-index ERIs are approximated by 3-index and 2-index ERIs using the auxiliary basis $\{P\}$:

$$(\alpha\beta|\gamma\delta) \approx \sum_{PR}^{N_{aux}} (\alpha\beta|P)[V^{-1}]_{PR} (R|\gamma\delta)$$

$$= \sum_Q^{N_{aux}} [\sum_P^{N_{aux}} [(\alpha\beta|P)[V^{-1/2}]_{PQ}][\sum_R^{N_{aux}} [V^{-1/2}]_{QR} (R|\gamma\delta)] \quad (1)$$

Where we have defined 4-, 3- and 2-index ERIs as

$$(\alpha\beta|\gamma\delta) = \iint dr_1 \, dr_2 \frac{\chi_\alpha(r_1)\chi_\beta(r_1)\chi_\gamma(r_2)\chi_\delta(r_2)}{r_{12}} \quad (2)$$

$$(\alpha\beta|P) = \iint dr_1 \, dr_2 \frac{\chi_\alpha(r_1)\chi_\beta(r_1)\chi_P(r_2)}{r_{12}} \quad (3)$$

$$V_{PQ} = (P|Q) = \iint dr_1 \, dr_2 \frac{\chi_P(r_1) \chi_Q(r_2)}{r_{12}} \tag{4}$$

Defining

$$K_{\alpha\beta}^Q \equiv \sum_P^{N_{aux}} (\alpha\beta|P) V_{PQ}^{-1/2} \tag{5}$$

we can rewrite the RI approximation as

$$(\alpha\beta|\gamma\delta) \approx \sum_Q^{N_{aux}} K_{\alpha\beta}^Q K_{\gamma\delta}^Q \tag{6}$$

Note that the formation of the RI matrix $K_{\alpha\beta}^Q$ in Eq. (5) scales as $O(N_{aux}^2 N_{AO}^2)$. Furthermore, the transformation of this 3-index matrix from AO basis to MO basis can be done in two steps,

$$K_{p\beta}^Q = \sum_\alpha^{N_{AO}} C_\alpha^p K_{\alpha\beta}^Q \tag{7}$$

$$K_{pq}^Q = \sum_\beta^{N_{AO}} C_\beta^q K_{p\beta}^Q \tag{8}$$

which scale as $O(N_{aux} N_{AO}^3)$. Here $C_\alpha^p$ and $C_\beta^q$ are the usual SCF MO coefficients. Since both $N_{aux}$ and $N_{AO}$ scale linearly with the system size $N$, the formal scaling of the above transformation is $O(N^5)$.

The stochastic realization of the RI approximation utilizes another set of stochastic orbitals $\{\theta^\xi\}$, $\xi = 1, 2, \ldots$, $N_s$ (with $N_s$ being the number of stochastic orbitals). All these stochastic orbitals are column arrays of length $N_{aux}$ with random elements $\pm 1$, (i.e. $\theta_A^\xi = \pm 1$). Thus, due to the central limit theorem, we have the following identity:

$$\langle \theta \otimes \theta \rangle_\xi = \frac{1}{N_s} \sum_{\xi=1}^{N_s} \theta^\xi \otimes (\theta^\xi)^T = \begin{pmatrix} \langle \theta_1 \theta_1 \rangle_\xi & \langle \theta_2 \theta_1 \rangle_\xi & \cdots & \langle \theta_1 \theta_{N_{aux}} \rangle_\xi \\ \langle \theta_2 \theta_1 \rangle_\xi & \langle \theta_2 \theta_2 \rangle_\xi & & \langle \theta_2 \theta_{N_{aux}} \rangle_\xi \\ \vdots & & \ddots & \vdots \\ \langle \theta_{N_{aux}} \theta_1 \rangle_\xi & \langle \theta_{N_{aux}} \theta_2 \rangle_\xi & \cdots & \langle \theta_{N_{aux}} \theta_{N_{aux}} \rangle_\xi \end{pmatrix} \approx I \tag{9}$$

In Eq. (9), since $\theta_A^\xi$ (and $\theta_B$) is a random choice of $\pm 1$, the diagonal matrix element denoted by $\langle \theta_A \theta_A \rangle_\xi$ always equals 1; the off-diagonal element denoted by $\langle \theta_A \theta_B \rangle_\xi$, however, converges to 0 when averaging over $N_s$ stochastic orbitals.

With the introduction of the stochastic resolution of identity, we can approximate 4-index ERIs using the stochastic orbitals in the following form:

$$(\alpha\beta|\gamma\delta) = \sum_{QS}^{N_{aux}} \sum_{PR}^{N_{aux}} (\alpha\beta|P) V_{PQ}^{-1/2} I_{QS} V_{SR}^{-1/2} (R|\gamma\delta)$$

$$\approx \sum_{QS}^{N_{aux}} \sum_{PR}^{N_{aux}} (\alpha\beta|P) V_{PQ}^{-1/2} \left(\langle \theta \otimes \theta^T \rangle_\xi\right)_{QS} (R|\gamma\delta) V_{SR}^{-1/2}$$

$$= \left\langle \left[\sum_P^{N_{aux}} (\alpha\beta|P) \sum_Q^{N_{aux}} \left(V_{PQ}^{-1/2} \theta_Q\right)\right] \left[\sum_R^{N_{aux}} (R|\gamma\delta) \sum_S^{N_{aux}} \left(V_{SR}^{-1/2} \theta_S^T\right)\right] \right\rangle_\xi \quad (10)$$

Similar to the RI case, we can define $R_{\alpha\beta}^\xi$ as

$$R_{\alpha\beta}^\xi = \sum_P^{N_{aux}} (\alpha\beta|P) \left[\sum_Q^{N_{aux}} \left(V_{PQ}^{-1/2} \theta_Q\right)\right] \quad (11)$$

such that we can rewrite the 4-index ERIs as

$$(\alpha\beta|\gamma\delta) \approx \frac{1}{N_s} \sum_{\xi=1}^{N_s} R_{\alpha\beta}^\xi R_{\gamma\delta}^\xi \equiv \left\langle R_{\alpha\beta}^\xi R_{\gamma\delta}^\xi \right\rangle_\xi \quad (12)$$

Note that the calculation of the $R_{\alpha\beta}^\xi$ matrix in Eq. (11) scales as $O(N_s N_{aux} N_{AO}^2)$. Furthermore, the transformation of the sRI matrix from AO basis to MO basis scales as $O(N_s N_{AO}^3)$,

$$R_{p\beta}^\xi = \sum_\alpha^{N_{AO}} C_\alpha^p R_{\alpha\beta}^\xi \quad (13)$$

$$R_{pq}^\xi = \sum_\beta^{N_{AO}} C_\beta^q R_{p\beta}^\xi \quad (14)$$

Thus the overall calculation of sRI matrix $R_{pq}^\xi$ formally scales as $O(N_s N^3)$. As shown previously (Ref. 19-27), when applying the sRI approach to intensive quantities, the number of stochastic orbitals $N_s$ being used is independent from the system size, such that the formation of the sRI matrix scales as $O(N^3)$. In particular, as shown below, we will apply the sRI approach to CC2 theory, where we can achieve an overall $O(N^3)$ scaling for ground state calculation. We will now introduce our sRI-CC2 theory in the following subsection.

### C. CC2 theory

In the CC2 model, the Hamiltonian $H$ undergoes a $T_1$-transformation

$$\hat{H} = exp(-T_1) H exp(T_1) \quad (15)$$

Here $T_1$ is the single excitation cluster operator. The singles and doubles equations which determine the amplitudes of a coupled cluster wavefunction can be written as

$$\Omega_{\mu_1} = \langle \mu_1 | \hat{H} + [\hat{H}, T_2] | HF \rangle = 0 \tag{16}$$

$$\Omega_{\mu_2} = \langle \mu_2 | \hat{H} + [F, T_2] | HF \rangle = 0 \tag{17}$$

where $\Omega_{\mu_1}$ and $\Omega_{\mu_2}$ are the single and double excitation vectors, $\mu_1$ and $\mu_2$ are the single and double excitation manifolds, and $|HF\rangle$ the Hartree-Fock reference state. In the CC2 theory, the vectors $\Omega_{\mu_1}$ and $\Omega_{\mu_2}$ can be expressed as

$$\Omega_{\mu_1} = \Omega_{ai} = \Omega_{ai}^G + \Omega_{ai}^H + \Omega_{ai}^I + \Omega_{ai}^J \tag{18}$$

$$\Omega_{\mu_2} = \Omega_{aibj} = \frac{1}{1 + \delta_{ij}\delta_{ab}} \left( \Omega_{aibj}^E + \Omega_{aibj}^F \right) \tag{19}$$

The terms in the above equations denote the different contributions to the CC equations, which can be written explicitly:

$$\Omega_{ai}^G = +\sum_{dlc} \hat{t}_{il}^{cd}(ld\hat{|}ac) \qquad \Omega_{ai}^H = -\sum_{dlk} \hat{t}_{kl}^{ad}(ld\hat{|}ki)$$

$$\Omega_{ai}^I = \sum_{dl} \hat{t}_{il}^{ad} \hat{F}_{ld} \qquad \Omega_{ai}^J = \hat{F}_{ai} \tag{20}\sim(23)$$

$$\Omega_{aibj}^E = (\varepsilon_i - \varepsilon_a + \varepsilon_j - \varepsilon_b) t_{ij}^{ab} \qquad \Omega_{aibj}^F = (ai\hat{|}bj) \tag{24}\sim(25)$$

In the above equations, $\hat{F}$ is the Fock matrix obtained from the $T_1$-transformed Hamiltonian. $(ai\hat{|}bj)$ is the transformed two-electron MO integrals given by:

$$(pq\hat{|}rs) = \sum_{\alpha\beta\gamma\delta} \Lambda_{\alpha p}^p \Lambda_{\beta q}^h \Lambda_{\gamma r}^p \Lambda_{\delta s}^h (\alpha\beta|\gamma\delta) \tag{26}$$

$$\Lambda^p = C(I - t_1^T) \tag{27}$$

$$\Lambda^h = C(I + t_1) \tag{28}$$

Here $t_1$ in the transformation matrices $\Lambda^p$ and $\Lambda^h$ for particle and hole operators is given by the auxiliary matrix comprised of singles cluster amplitudes $\{t_i^a\}$:

$$t_1 = \begin{pmatrix} 0 & 0 \\ \{t_i^a\} & 0 \end{pmatrix} \tag{29}$$

Finally, $\hat{t}_{ij}^{ab}$ in the CC2 vector equations is given by

$$t_{ij}^{ab} = \frac{(ai\hat{|}bj)}{\varepsilon_i - \varepsilon_a + \varepsilon_j - \varepsilon_b} \tag{30}$$

$$\hat{t}_{ij}^{ab} = (1 + \delta_{ij}\delta_{ab})(2\, t_{ij}^{ab} - t_{ij}^{ba}) \tag{31}$$

To solve the CC2 vector equations, we start from a set of trial $\{t_i^a\}$. The double excitation amplitudes $t_{ij}^{ab}$ as well as their transformed ones can be calculated as intermediates, such that we can determine $\Omega_{ai}$. The DIIS algorithm helps us to update $\{t_i^a\}$, such that we can minimize $\Omega_{ai}$. With the converged amplitudes, we can calculate the total correlation energy:

$$E_{corr} = \sum_{aibj} t_{ij}^{ab} [2(ai|bj) - (bi|aj)] \tag{32}$$

Note that the CC2 theory scales as $O(N^5)$ with the system size. The RI approach decreases the prefactor of the CC2 calculations, yet the overall scaling of RI-CC2 is still $O(N^5)$, prohibiting the usage of CC2 theory to large systems. Below, we introduce the algorithm of sRI-CC2 theory, which can formally bring down the scaling to $O(N^3)$.

### D. Algorithm for sRI-CC2

The main steps in the scheme of the algorithm for sRI-CC2 are summarized as follows (Here we employ Einstein Summation rule):

1. Construct stochastic orbitals $R_{ai}^{\xi}$ from Eq. (14) and initialize all the single excitation amplitudes $t_i^a = 0$.

2. Calculate $\Lambda_{\alpha p}^p$ and $\Lambda_{\alpha p}^h$ as well as $\hat{R}_{pq}^{\xi}$

$$\hat{R}_{p\beta}^{\xi} = \sum_{\alpha}^{N_{AO}} \Lambda_{\alpha p}^p R_{\alpha\beta}^{\xi} \qquad \hat{R}_{pq}^{\xi} = \sum_{\beta q}^{N_{AO}} \Lambda_{\beta q}^h \hat{R}_{p\beta}^{\xi} \tag{33}\sim(34)$$

3. Calculate $\Omega_1 = (\varepsilon_a - \varepsilon_i)\, t_i^a$ $\tag{35}$

4. Construct $\hat{F}_{ia} = \left\langle 2R_{ia}^{\xi}(R_{kc}^{\xi} t_c^k) - R_{ja}^{\xi}(R_{ib}^{\xi} t_b^j) \right\rangle_{\xi}$ $\tag{36}$

5. Calculate $\Omega_2 = \Omega_{ai}^I = \hat{t}_{ij}^{ab} \hat{F}_{jb}$. With the help of sRI and Laplace transformation, we can rewrite $\Omega_2$ as follows:

$$\Omega_2 = \Omega_{ai}^I = \hat{t}_{ij}^{ab} \hat{F}_{jb} = \left( \frac{2\langle \hat{R}_{ai}^{\xi} \hat{R}_{bj}^{\xi} \rangle_{\xi}}{\varepsilon_i - \varepsilon_a + \varepsilon_j - \varepsilon_b} - \frac{\langle \hat{R}_{aj}^{\xi} \hat{R}_{bi}^{\xi} \rangle_{\xi}}{\varepsilon_i - \varepsilon_a + \varepsilon_j - \varepsilon_b} + \frac{\langle \hat{R}_{ai}^{\xi} \hat{R}_{ai}^{\xi} \rangle_{\xi}}{2(\varepsilon_i - \varepsilon_a)} \right) \hat{F}_{jb}$$

$$= -2 \int_0^{\infty} \left\langle [\hat{R}_{bj}^{\xi} e^{(\varepsilon_j - \varepsilon_b)t} \hat{F}_{jb}] \hat{R}_{ai}^{\xi} e^{(\varepsilon_i - \varepsilon_a)t} \right\rangle_{\xi} dt + \int_0^{\infty} \left\langle [\hat{R}_{bi}^{\xi} e^{(-\varepsilon_b)t} \hat{F}_{jb}] \hat{R}_{aj}^{\xi} e^{(\varepsilon_i - \varepsilon_a + \varepsilon_j)t} \right\rangle_{\xi} dt$$

$$+ \frac{\langle \hat{R}_{ai}^{\xi} \hat{R}_{ai}^{\xi} \rangle_{\xi}}{2(\varepsilon_i - \varepsilon_a)} \hat{F}_{ia} \tag{37}$$

Note that the evaluation of the first two terms in the above equation requires integration over time. This is done with Gaussian quadrature, such that computational scaling of these two terms are $O(N_s N_t N^2)$ and $O(N_s N_t N^3)$ respectively, where $N_t$ is the number of quadrature points. Obviously, the last term is the above equation scales as $O(N_s N^2)$. Since $N_t$ is independent from the system size, the overall scaling of $\Omega_2$ is $O(N^3)$.

6. Calculate $\Omega_3$ and $\Omega_4$. Since these two terms are very similar, we only demonstrate how to determine $\Omega_3$ in details. In the RI approach, $\Omega_3$ can be expressed as

$$\Omega_3 = \Lambda_{\alpha a}^p(\alpha\beta|P) V_{PQ}^{-1/2} \Lambda_{\beta a}^h \hat{t}_{ij}^{ab} K_{jb}^Q = \hat{t}_{ij}^{ab} \hat{K}_{aa}^Q K_{jb}^Q \tag{38}$$

In the sRI approach, we use two sets of independent stochastic orbitals denoted by $\xi$ and $\xi'$ to approximate the ERIs, such that

$$\Omega_3 = \left( \frac{2\langle \hat{R}_{ai}^\xi \hat{R}_{bj}^\xi \rangle_\xi}{\varepsilon_i - \varepsilon_a + \varepsilon_j - \varepsilon_b} - \frac{\langle \hat{R}_{aj}^\xi \hat{R}_{bi}^\xi \rangle_\xi}{\varepsilon_i - \varepsilon_a + \varepsilon_j - \varepsilon_b} + \frac{\langle \hat{R}_{ai}^\xi \hat{R}_{ai}^\xi \rangle_\xi}{2(\varepsilon_i - \varepsilon_a)} \right) \langle \hat{R}_{aa}^{\xi'} R_{jb}^{\xi'} \rangle_{\xi'}$$

$$= -2 \int_0^\infty \langle [\hat{R}_{bj}^\xi e^{(\varepsilon_j-\varepsilon_b)t} R_{jb}^{\xi'}] \hat{R}_{ai}^\xi e^{(\varepsilon_i-\varepsilon_a)t} \hat{R}_{aa}^{\xi'} \rangle_{\xi\xi'} dt + \int_0^\infty \langle [\hat{R}_{bi}^\xi e^{(-\varepsilon_b)t} R_{jb}^{\xi'}] \hat{R}_{aj}^\xi e^{(\varepsilon_i-\varepsilon_a+\varepsilon_j)t} \hat{R}_{aa}^{\xi'} \rangle_{\xi\xi'} dt$$

$$+ \left\langle \frac{\hat{R}_{ai}^\xi \hat{R}_{ai}^\xi}{2(\varepsilon_i - \varepsilon_a)} R_{ia}^{\xi'} \hat{R}_{aa}^{\xi'} \right\rangle_{\xi\xi'} \tag{39}$$

Here, again we have used Laplace transformation in the last line of the above equation. The overall scaling of $\Omega_3$ is $O(N^3)$. Evaluation of $\Omega_4$ can be done in the similar manner (See Appendix A).

7. Calculate $\Omega_5$ and $\Omega_6$. The last two parts of $\Omega_{ai}$ are easy to obtain:

$$\Omega_5 = 2 \langle (t_c^k R_{kc}^\xi) \hat{R}_{ai}^\xi \rangle_\xi \tag{40}$$

$$\Omega_6 = -\langle (t_k^c R_{ki}^\xi) \hat{R}_{ac}^\xi \rangle_\xi \tag{41}$$

8. Combine $\Omega_{ai} = \Omega_1 + \Omega_2 + \Omega_3 + \Omega_4 + \Omega_5 + \Omega_6$. Update the $\{t_i^a\}$ as follows:

$$t_i^a(updated) = t_i^a(old) - \frac{\Omega_{ai}}{\varepsilon_a - \varepsilon_i} \tag{42}$$

9. The steps 2-8 are iterated until self-consistency is reached.

In the original RI-CC2 algorithm, the limiting computational step is determining $\Omega_3$ and $\Omega_4$, which requires computing $t_{ij}^{ab}$ as intermediate, such that the overall scaling of RI-CC2 is $O(N^5)$. With sRI approach, no $t_{ij}^{ab}$ is computed directly, and we bring down the scaling to $O(N^3)$. Finally, the CC2 ground-state energy in Eq. (32) is calculated in a similar manner using sRI and Laplace transformation:

$$E_{corr} = \sum_{aibj} t_{ij}^{ab} [2(ai|bj) - (bi|aj)] = \sum_{aibj} \frac{2(ai|bj) - (bi|aj)}{\varepsilon_i - \varepsilon_a + \varepsilon_j - \varepsilon_b}(ai\hat{|}bj)$$

$$= \sum_{aibj} \frac{2\langle R_{ai}^{\xi}R_{bj}^{\xi}\rangle_{\xi} - \langle R_{bi}^{\xi}R_{aj}^{\xi}\rangle_{\xi}}{\varepsilon_i - \varepsilon_a + \varepsilon_j - \varepsilon_b}\langle \hat{R}_{ai}^{\xi'}\hat{R}_{bj}^{\xi'}\rangle_{\xi'} = \left\langle \sum_{aibj} \frac{2R_{ai}^{\xi}R_{bj}^{\xi} - R_{bi}^{\xi}R_{aj}^{\xi}}{\varepsilon_i - \varepsilon_a + \varepsilon_j - \varepsilon_b}\hat{R}_{ai}^{\xi'}\hat{R}_{bj}^{\xi'}\right\rangle_{\xi\xi'}$$

$$= -\int_0^{\infty} \left\langle \sum_{aibj} [2(R_{ai}^{\xi}\hat{R}_{ai}^{\xi'})(R_{bj}^{\xi}\hat{R}_{bj}^{\xi'}) - (R_{aj}^{\xi}\hat{R}_{ai}^{\xi'})(R_{bi}^{\xi}\hat{R}_{bj}^{\xi'})]e^{(\varepsilon_i - \varepsilon_a + \varepsilon_j - \varepsilon_b)t}\right\rangle_{\xi\xi'} dt$$

$$= -\int_0^{\infty} \langle 2A(t)^2 - Tr(E(t)^2)\rangle_{\xi\xi'} dt \quad (43)$$

where

$$A(t) = \sum_{ai} e^{(\varepsilon_i - \varepsilon_a)t} R_{ai}^{\xi}\hat{R}_{ai}^{\xi'} \quad (44)$$

$$E(t)_{ij} = \sum_a e^{(\varepsilon_i - \varepsilon_a)t} R_{aj}^{\xi}\hat{R}_{ai}^{\xi'} \quad (45)$$

Still, the calculation of the ground state energy scales as $O(N_s N_t N^3)$, seen as $O(N^3)$. Thus the overall scaling of sRI-CC2 is $O(N^3)$. Below, we apply our sRI-CC2 approach to a series of molecules. In particular, we demonstrate the scaling of sRI-CC2 and analyze the stochastic errors of the ground state energy.

## III. RESULTS AND DISCUSSION

We now apply the sRI-CC2 algorithm to a series of molecular systems, including a hydrogen dimer chain $H_n$, where n is the number of hydrogen atoms, ranging from 10 to 400. In each dimer, the distance between two hydrogen atoms is 0.74 Å, and the distance between two hydrogen atoms of each adjacent dimer is 1.26 Å. We test the performance of sRI-CC2 against RI-CC2 in Q-Chem package; in particular we focus on the scaling, the stochastic error, and the influence of prefactor $N_s$ on the ground state energy especially on large-size systems. In addition, a list of small molecules as well as n-alkanes $C_nH_{2n+2}$ are also selected to test the performance of the sRI-CC2 approach.

### A. Correlation energy

In Table II, we show the correlation energy per electron for a list of different hydrogen dimer chains ($H_{10}$, $H_{80}$, $H_{200}$, $H_{400}$) using a sto-3g basis. $N_s$ = 200, 400 and 800 stochastic orbitals are used for ground state energy calculations and the final results are obtained by averaging over 10 independent runs. Notice that with increasing number of stochastic orbitals, the standard deviation of the correlation energy decreases. We further compare the results from sRI-CC2 calculations with 400 stochastic orbitals against RI-CC2 results in Figure 1, where we see perfect agreement.

Note that we can easily go to $H_{400}$ (and beyond) in sRI-CC2 calculations, whereas such a large system is very difficult for RI-CC2 calculation. Note also that for a weakly correlated system such as the hydrogen dimer chains, the correlation energy shows a linear dependence with the number of dimer chains.

TABLE II. Correlation energy per electron (in mEh). Here $N_H$ is the number of hydrogen in the corresponding dimer chain, the same value as the number of correlated electrons $N_e$. The standard deviation $\sigma$ is calculated from 10 independent samples.

| $N_H$ | $E_{corr}$ per electron (mEh) | | | std deviation per electron (mEh) | | |
|---|---|---|---|---|---|---|
| | $N_S = 200$ | $N_S = 400$ | $N_S = 800$ | $N_S = 200$ | $N_S = 400$ | $N_S = 800$ |
| $H_{10}$ | -5.954 | -6.630 | -6.777 | 0.466 | 0.581 | 0.522 |
| $H_{80}$ | -6.889 | -6.999 | -7.109 | 1.071 | 0.866 | 0.608 |
| $H_{200}$ | -7.401 | -7.051 | -7.033 | 1.088 | 0.970 | 0.513 |
| $H_{400}$ | -7.690 | -6.989 | -7.252 | 0.861 | 1.107 | 0.735 |

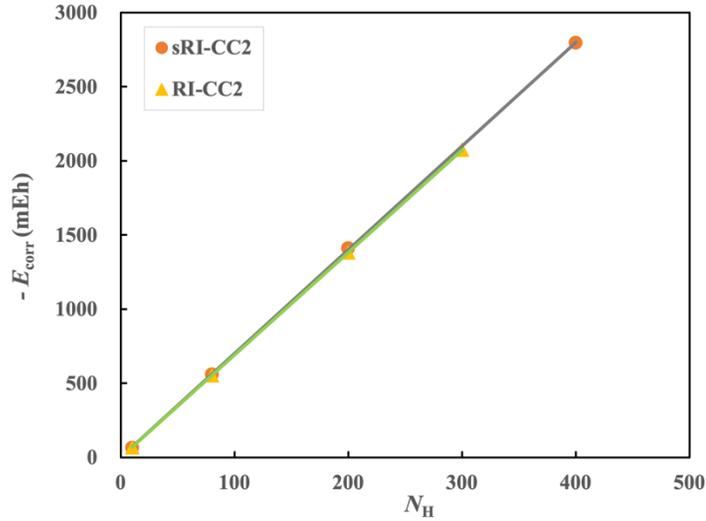

FIG. 1. Correlation energy for sRI-CC2 as a function of the number of hydrogen atoms with $N_s = 400$. The difference compared with the RI-CC2 can be approximately neglected, which shows the high performance in accuracy.

In Figure 2, we plot the correlation energy per electron for hydrogen dimer chains from sRI-CC2 and RI-CC2. Again, we use 400 stochastic orbitals in our sRI-CC2 calculations, and the results are averaged from 10 independent runs. Furthermore, the error bars indicate the standard deviations of these 10 independent runs. We see that the sRI-CC2 results agree with RI-CC2 results within the error bars, which shows the validity of sRI-CC2 approximation. Furthermore, we observe that the statistical error bar for energy per electron does not increase with the system size and almost keeps unchanged. This observation indicates that we do not need to increase the number of stochastic orbitals when calculating intensive quantities for larger systems to achieve the same accuracy as smaller ones. Thus the prefactor $N_s$ is fixed for different system size, such that the overall scaling of sRI-CC2 calculation is $O(N^3)$. Similar conclusion has been reported in sRI-MP2[21] and sRI-GF2[26] approaches.

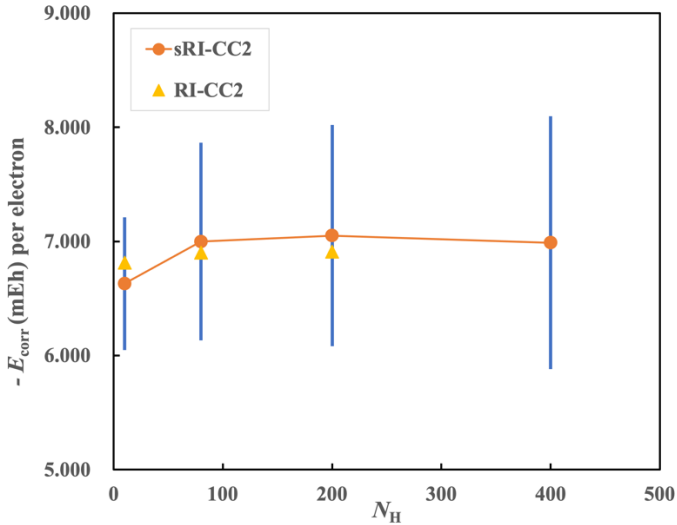

FIG. 2. Correlation energy per electron as a function of the number of hydrogen atoms with $N_s = 400$. The error bar is from the standard deviation calculated from 10 different seeds.

**B. Error assessment**

In Figure 3, we further analyze the statistical error of the correlation energy per electron as a function of number of stochastic orbitals ($N_s$ = 200, 400, and 800) for a set of hydrogen dimer chains. Again, we see the deceases of the error bars when increasing the number of stochastic orbitals. Due to the central limit theorem, the error bars decrease with the number of stochastic orbitals as $1/\sqrt{N_s}$, which is consistent with our results. In addition, we do not see obvious linear dependency of the averaged correlation energy on the number of stochastic orbitals, which indicates that we do not have a bias in our calculations. Such a bias is reported previously in sGF2[25]. No such a bias is reported in sRI-GF2 results[26].

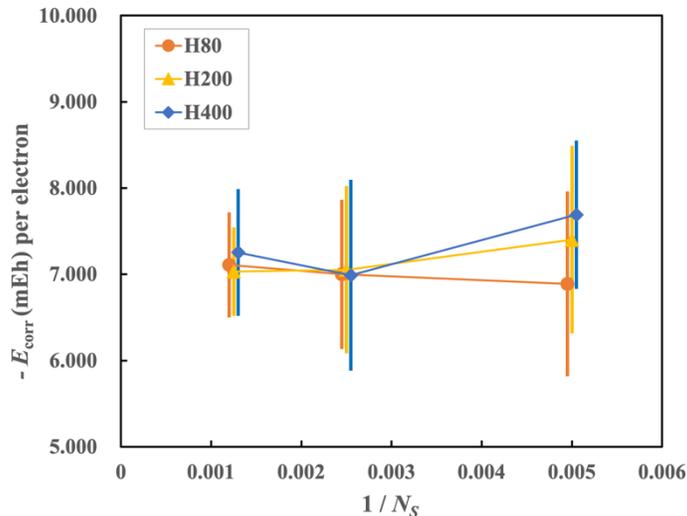

FIG. 3. Correlation energy per electron as a function of the reciprocal of the number of stochastic orbitals with $N_s$ = 200, 400 and 800. 10 independent calculations of sRI-CC2 are used to estimate the error bar as well.

## C. Scaling

We now turn our attention to the computational cost of the sRI-CC2 (as well as RI-CC2 in the Q-Chem program package) calculations. All the calculations are performed in an AMD EPYC 7502 (2.5GHz) node with 64 computational cores. Again, we use the hydrogen dimer chains as our test set.

In Figure 4, we show the computational time of sRI-CC2 and RI-CC2 for the hydrogen dimer chains. We use $N_s$ = 400 orbitals in our sRI-CC2 calculations. We observe experimental scaling of $O(N_H^{3.74})$ for RI-CC2 approach in Q-Chem. The experimental scaling of the sRI-CC2 approach is $O(N_H^{2.71})$ from fitting, slightly better than the theoretical scaling $O(N^3)$. Our ongoing attempt to transplant our sRI-CC2 to Q-Chem would further lower this scaling. Moreover, the sRI-CC2 approach can be easily parallelized over stochastic orbitals using MPI or OpenMP. We expect that the sRI-CC2 approach can be easily used to calculate systems with 1000 electrons or more.

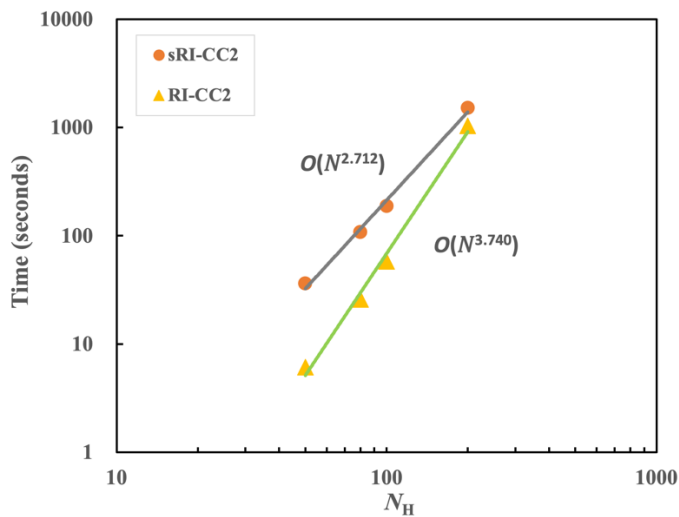

FIG. 4. CPU time as a function of the number of hydrogen atoms with $N_s$ = 400. The experimental scaling of the RI-CC2 method is $O(N_H^{3.74})$, and the experimental scaling of the sRI-CC2 is $O(N_H^{2.71})$.

### D. Supplementary calculation

To test our sRI-CC2 algorithm besides $H_n$, we apply it to a set of other molecules. The experimental geometries are taken from the reference[29] and are included in the Supporting Information. In Figures 5 and 6, we selectively show the results of correlation energies per electron of both algorithms and standard deviations with $N_S = 400$ and cc-pVDZ/aug-cc-pVDZ basis. These two figures are very similar and all the data points of RI-CC2 are located on the error bars. Although these selected systems are quite different, the errors between RI and sRI are acceptable (see Table III in Appendix B).

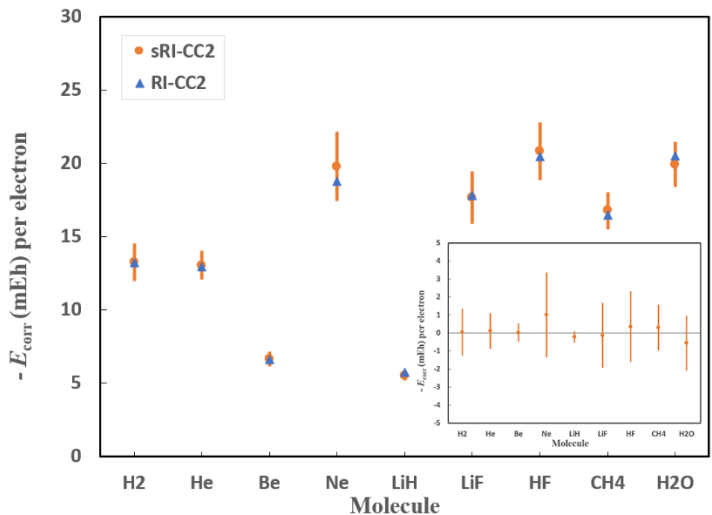

FIG. 5. Correlation energy per electron for 9 different molecular systems with cc-pVDZ basis set and $N_S = 400$. The error bar is from the standard deviation calculated from 10 different seeds (inset: Take the RI-CC2 energy as the benchmark).

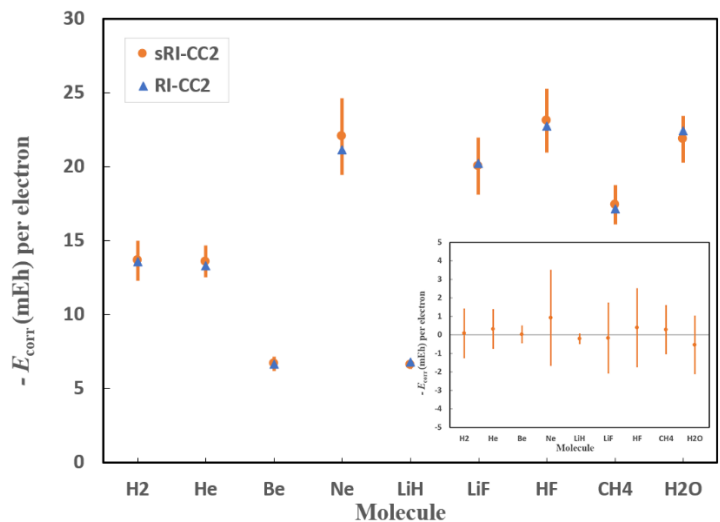

FIG. 6. Correlation energy per electron for 9 different molecular systems with aug-cc-pVDZ basis set and $N_s = 400$. The error bar is from the standard deviation calculated from 10 different seeds (inset: Take the RI-CC2 energy as the benchmark).

For further exploration, we perform calculations on a series of n-alkanes $C_nH_{2n+2}$ with n = 1~5 using 400 stochastic orbitals. The experimental geometries are taken from the references[29-31]. This time we employ cc-pvXZ and aug-cc-pvXZ (X = D, T) basis sets and all the data are listed in Table IV of Appendix B. Although we adopt larger basis sets this time, the standard deviations remain small and hardly changed with the system size. Similarly, the results of cc-pVDZ and cc-pVTZ basis sets are presented in Figures 7 and 8.

In summary, the assessment in this subsection again highlights the wide application scope of our sRI-CC2 method for various molecular systems and basis sets.

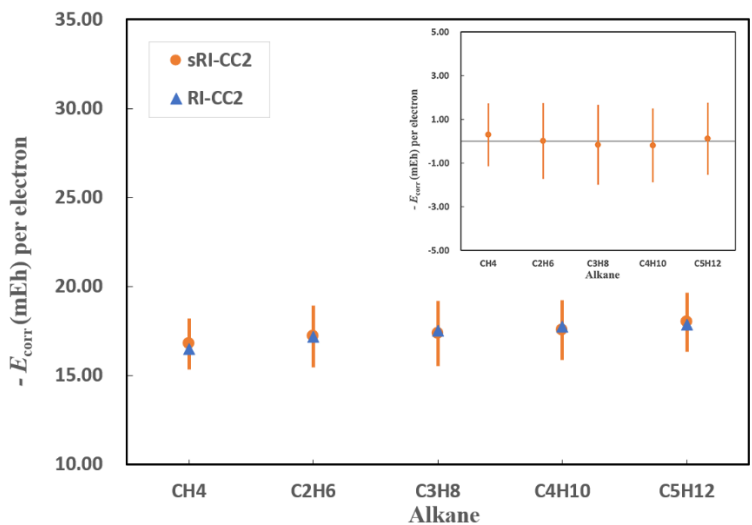

FIG. 7. Correlation energy per electron for n-alkanes $C_nH_{2n+2}$, n ranging from 1~5, with cc-pVDZ basis set and $N_s = 400$. The error bar is from the standard deviation calculated from 10 different seeds (inset: Take the RI-CC2 energy as the benchmark).

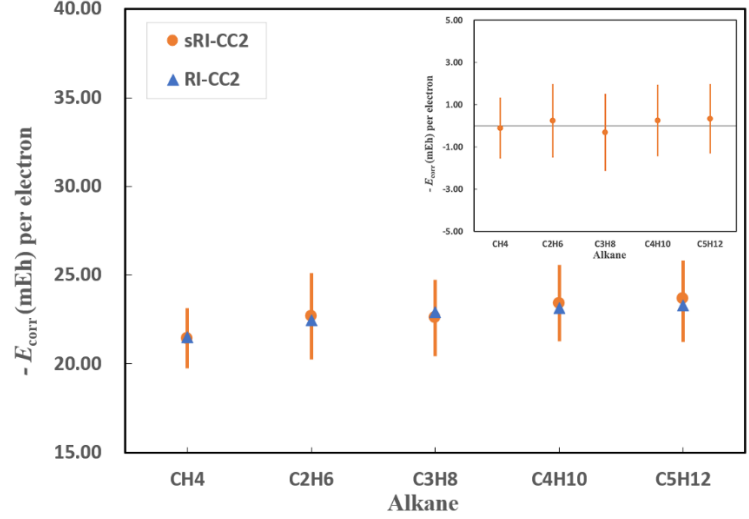

FIG. 8. Correlation energy per electron for n-alkanes $C_nH_{2n+2}$, n ranging from 1~5, with cc-pVTZ basis set and $N_s$ = 400. The error bar is from the standard deviation calculated from 10 different seeds (inset: Take the RI-CC2 energy as the benchmark).

## IV. CONCLUSIONS

We present a stochastic approach of the RI approximation to the CC2 model to calculate the ground state energy. Using this stochastic orbitals method, the 4-index ERIs in the algorithm for RI-CC2 are decoupled and the overall scaling of the sRI-CC2 is reduced from $O(N^5)$ to $O(N^3)$. Within the range of hydrogen dimer chains we test, the results of sRI in terms of energy accuracy show high performance compared with RI-CC2 and the observed scaling is $O(N_H^{2.71})$ from curve fitting. This sRI-CC2 approach provides a cheap and practical alternative for the ground-state energy calculations for different molecules and basis sets, especially for large systems. Future work on how to apply sRI to the calculation of excited-state energy of CC2 model is in progress, which we may pay more attention to.

## ACKNOWLEDGEMENT

We acknowledge the startup funding from Westlake University. We thank Joonho Lee for helpful discussions. We also acknowledge high performance computing (HPC) service at Westlake University. C. Z. thanks Jian Zhu for helpful suggestions on program development.

## APPENDIX A: THE EXPLICIT EXPRESSION FOR A PARTITION OF EXCITATION VECTOR

$\Omega_4$ is evaluated in the similar manner with $\Omega_3$:

$$\Omega_4 = -Y_{ak}^P K_{ki}^P = -\hat{t}_{kj}^{ab} K_{jb}^P K_{ki}^P = -\left( \frac{2\langle \hat{R}_{ak}^\xi \hat{R}_{bj}^\xi \rangle_\xi}{\varepsilon_k - \varepsilon_a + \varepsilon_j - \varepsilon_b} - \frac{\langle \hat{R}_{aj}^\xi \hat{R}_{bk}^\xi \rangle_\xi}{\varepsilon_k - \varepsilon_a + \varepsilon_j - \varepsilon_b} + \frac{\langle \hat{R}_{ak}^\xi \hat{R}_{ak}^\xi \rangle_\xi}{2(\varepsilon_k - \varepsilon_a)} \right) \langle R_{jb}^{\xi'} R_{ki}^{\xi'} \rangle_{\xi'},$$

$$= 2\int_0^\infty \left\langle \left[\hat{R}_{bj}^\xi e^{(\varepsilon_j-\varepsilon_b)t} R_{jb}^{\xi'}\right] \hat{R}_{ak}^\xi e^{(\varepsilon_k-\varepsilon_a)t} R_{ki}^{\xi'}\right\rangle_{\xi\xi'} dt - \int_0^\infty \left\langle \left[\hat{R}_{bk}^\xi e^{(-\varepsilon_b)t} R_{jb}^{\xi'}\right] \hat{R}_{aj}^\xi e^{(\varepsilon_k-\varepsilon_a+\varepsilon_j)t} R_{ki}^{\xi'}\right\rangle_{\xi\xi'} dt$$

$$- \left\langle \frac{\hat{R}_{ak}^\xi \hat{R}_{ak}^\xi}{2(\varepsilon_k-\varepsilon_a)} R_{ka}^{\xi'} R_{ki}^{\xi'}\right\rangle_{\xi\xi'}, \tag{46}$$

## APPENDIX B: SUPPLEMENTARY DATA

Results for 9 molecules with $N_s = 400$ and cc-pVDZ/aug-cc-pVDZ basis sets are displayed in TABLE III. And data for n-alkanes $C_nH_{2n+2}$, n ranging from 1~5, using 400 stochastic orbitals and cc-pvXZ/aug-cc-pvXZ (X = D, T) basis sets, are shown in TABLE IV. The values of all errors between RI and sRI are acceptable and within the standard deviations.

TABLE III. RI-CC2 and sRI-CC2 results for 9 different systems. Here $N_e$ is the number of correlated electrons. The correlation energies per electron, errors and the standard deviations per electron of 10 sRI runs are in mEh.

| $N_e$ | molecule | cc-pVDZ | | | | aug-cc-pVDZ | | | |
|---|---|---|---|---|---|---|---|---|---|
| | | RI | sRI | error | std deviation | RI | sRI | error | std deviation |
| 2 | H$_2$ | -13.213 | -13.258 | 0.044 | 1.295 | -13.576 | -13.662 | 0.086 | 1.344 |
| 2 | He | -12.915 | -13.029 | 0.115 | 0.986 | -13.289 | -13.605 | 0.316 | 1.073 |
| 4 | Be | -6.621 | -6.640 | 0.019 | 0.515 | -6.668 | -6.692 | 0.024 | 0.482 |
| 10 | Ne | -18.779 | -19.783 | 1.005 | 2.358 | -21.145 | -22.060 | 0.915 | 2.595 |
| 4 | LiH | -5.721 | -5.505 | 0.216 | 0.318 | -6.817 | -6.609 | 0.209 | 0.285 |
| 12 | LiF | -17.805 | -17.667 | 0.139 | 1.796 | -20.215 | -20.043 | 0.172 | 1.911 |
| 10 | HF | -20.452 | -20.806 | 0.354 | 1.968 | -22.737 | -23.126 | 0.389 | 2.140 |
| 10 | CH$_4$ | -16.461 | -16.765 | 0.303 | 1.273 | -17.167 | -17.449 | 0.282 | 1.325 |
| 10 | H$_2$O | -20.481 | -19.922 | 0.559 | 1.524 | -22.412 | -21.874 | 0.539 | 1.587 |

TABLE IV. RI-CC2 and sRI-CC2 results for of n-alkanes $C_nH_{2n+2}$ with n = 1~5. Here $N_e$ is the number of correlated electrons. The correlation energies per electron of RI and sRI, errors and the standard deviations per electron of 10 sRI runs are in mEh.

| $N_e$ | molecule | cc-pVDZ | | | | aug-cc-pVDZ | | | |
|---|---|---|---|---|---|---|---|---|---|
| | | RI | sRI | error | std deviation | RI | sRI | error | std deviation |
| 10 | $CH_4$ | -16.461 | -16.765 | 0.303 | 1.443 | -17.167 | -17.449 | 0.282 | 1.475 |
| 18 | $C_2H_6$ | -17.173 | -17.191 | 0.017 | 1.739 | -17.861 | -17.906 | 0.045 | 1.823 |
| 26 | $C_3H_8$ | -17.511 | -17.347 | 0.164 | 1.823 | -18.218 | -18.058 | 0.159 | 1.919 |
| 34 | $C_4H_{10}$ | -17.737 | -17.546 | 0.190 | 1.691 | -18.465 | -18.365 | 0.100 | 1.744 |
| 42 | $C_5H_{12}$ | -17.852 | -17.988 | 0.136 | 1.643 | -18.595 | -18.767 | 0.173 | 1.761 |

| $N_e$ | molecules | cc-pVTZ | | | | aug-cc-pVTZ | | | |
|---|---|---|---|---|---|---|---|---|---|
| | | RI | sRI | error | std deviation | RI | sRI | error | std deviation |
| 10 | $CH_4$ | -21.521 | -21.428 | 0.093 | 1.695 | -21.990 | -21.858 | 0.132 | 1.745 |
| 18 | $C_2H_6$ | -22.440 | -22.696 | 0.256 | 2.439 | -23.003 | -23.294 | 0.292 | 2.516 |
| 26 | $C_3H_8$ | -22.892 | -22.590 | 0.302 | 2.151 | -23.502 | -23.241 | 0.261 | 2.152 |
| 34 | $C_4H_{10}$ | -23.149 | -23.409 | 0.260 | 2.155 | -23.800 | -24.176 | 0.376 | 2.210 |
| 42 | $C_5H_{12}$ | -23.304 | -23.659 | 0.355 | 2.429 | -23.975 | -24.366 | 0.391 | 2.515 |

**APPENDIX C: DISTRIBUTION OF SRI-CC2 ENERGIES**

To test whether the sRI-CC2 energy results conform to a certain distribution, we apply the calculation scheme to the hydrogen dimer chain $H_{20}$ with $N_S = 800$. In other words, 800 randomly selected sampling seeds are used to obtain the ground-state energies respectively. With the average and standard derivation of these 800 runs, we plot a normal distribution curve in orange in Figure 5. And the real distribution of samples is plotted in the same figure in blue, well coinciding with the orange curve. Therefore, the results of sRI-CC2 energy with different sampling seeds basically meet the normal distribution.

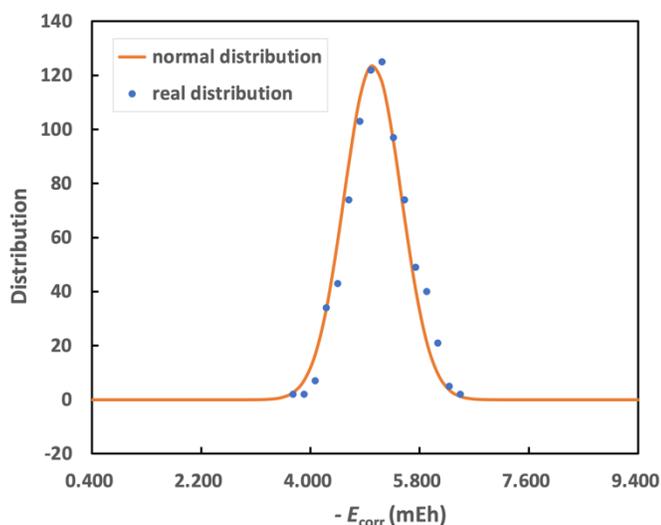

FIG. 9. Distribution of the sRI-CC2 ground state energy of the $H_{20}$ hydrogen dimer chain with $N_S = 800$. The curve fits well with the normal distribution.